# Growth-Induced In-Plane Uniaxial Anisotropy in $V_2O_3$/Ni Films

Dustin A. Gilbert,[1,2,*] Juan Gabriel Ramírez,[3] T. Saerbeck,[4] J. Trastoy,[5] Ivan K. Schuller,[5] Kai Liu,[1,*] and J. de la Venta[6,*]

[1]*Physics Department, University of California, Davis, California 95616*

[2]*NIST Center for Neutron Research, Gaithersburg, Maryland, 20899*

[3]*Department of Physics, Universidad de los Andes, Bogotá 111711, Colombia*

[4]*Institut Laue-Langevin, 71 Av. Des Martyrs, CS 20156, 38042 Grenoble cedex 9, France*

[5]*Physics Department, University of California, San Diego, California 92093*

[6]*Physics Department, Colorado State University, Fort Collins, Colorado 80523*

* Correspondence to dustin.gilbert@nist.gov (D.A.G.), kailiu@ucdavis.edu (K.L.), jose.de_la_venta@colostate.edu (J.D.L.V.).

**Abstract**

We report on a strain-induced and temperature dependent uniaxial anisotropy in $V_2O_3$/Ni hybrid thin films, manifested through the interfacial strain and sample microstructure, and its consequences on the angular dependent magnetization reversal. X-ray diffraction and reciprocal space maps identify the in-plane crystalline axes of the $V_2O_3$; atomic force and scanning electron microscopy reveal oriented rips in the film microstructure. Quasi-static magnetometry and dynamic ferromagnetic resonance measurements identify a uniaxial magnetic easy axis along the rips. Comparison with films grown on sapphire without rips shows a combined contribution from strain and microstructure in the $V_2O_3$/Ni films. Magnetization reversal characteristics captured by angular-dependent first order reversal curve measurements indicate a strong domain wall pinning along the direction orthogonal to the rips, inducing an angular-dependent change in the reversal mechanism. The resultant anisotropy is tunable with temperature and is most pronounced at room temperature, which is beneficial for potential device applications.

**Introduction**

Proximity effects in heterostructures are of keen interest in magnetism due to both emergent phenomena[1, 2, 3, 4] and potential applications, e.g., in advanced data storage and sensor technologies[5, 6, 7, 8, 9]. Interface coupling of magnetically ordered structures can induce behaviors such as exchange bias[10, 11, 12] and exchange spring[5, 13] effects. Recent studies have demonstrated strain-induced proximity effects in vanadium oxide ($V_2O_3$, $VO_2$)/Ni bilayer films[14, 15]. In particular, $V_2O_3$ undergoes a simultaneous rhombohedral to monoclinic (structural), paramagnetic to antiferromagnetic (magnetic), and metal to insulator (electronic) phase transition as the temperature is decreased below 160 K[16]. Strain induced in the Ni film by the $V_2O_3$ structural phase transition (SPT)[17] causes drastic changes in its magnetic properties, including a spike in the coercivity ($H_C$) midway through the $V_2O_3$ phase transition. Interestingly, the observed magnetic properties were highly dependent on the sample microstructure along the out-of-plane direction - specifically the interface roughness - although the strain was expected to be predominantly in the film plane[14, 15]. An interface roughness of 1.5 nm suppressed the coercivity spike, while a general increase in $H_C$ due to an isotropic strain persisted. This emphasizes the sensitivity of the $V_2O_3$/ Ni system to its microstructure. With the current interest in strain control of materials, including magnetic materials[18] and multiferroics,[19, 20, 21] among others,[22, 23] understanding the intricacies of strain and its potential interplay with the microstructure is of crucial importance. This interest, and the known sensitivity of the ($V_2O_3$, $VO_2$)/ Ni system, motivates an investigation on how the magnetic properties are impacted by the strain through the in-plane microstructure.

Here we present a detailed angular and temperature ($T$) dependent study of the magnetic characteristics of $V_2O_3$/Ni thin films. We observe long-range oriented terraces and long voids ("rips") in the microstructure, which induce a uniaxial magnetic anisotropy in the Ni film. Ferromagnetic resonance (FMR) shows a uniaxial enhancement of the in-plane field, and damping of the magnon excitation. Comparison measurements performed on sapphire/Ni samples without rips show similar uniaxial anisotropy, but without the magnon damping, suggesting that both the strain and microstructure are important in the observed properties of $V_2O_3$/Ni films. The quasi-static magnetization reversal was investigated with first order reversal curve (FORC) measurements. FORC diagrams show a different reversal mechanism when the field is applied parallel or orthogonal to the rips. Measurements performed at 45° relative to the rips show both reversal mechanisms, tunable by the measurement temperature. The FORC studies lead to understanding of the domain nucleation, propagation, and pinning behavior resulted from the microstructure and strain. The presence of oriented, in-plane magnetic anisotropy, emergent without patterning or magneto-crystalline considerations, offers new opportunities for scalable nanomagnetic devices.

**Results**

*Structure analysis.*

Films of $V_2O_3$ (100 nm)/Ni (10 nm)/Cu cap (7 nm) and Ni (10 nm)/Nb cap (6 nm) were grown on r-cut (102) sapphire substrates, as described in Methods. The XRD patterns, shown in Fig. 1, reveal a rhombohedral corundum structure in the as-grown $V_2O_3$ film with an (012) orientation. An out-of-plane lattice parameter of 3.658 Å is found, which agrees well with the expected corundum lattice constants of bulk $V_2O_3$. Ni is in its face centered cubic structure with a (111) texture and a lattice constant a 3.604±0.001 Å, close to reported bulk values. We assume polycrystalline growth of the Ni, but a hypothetical epitaxial relationship of $V_2O_3(012)\|Ni(111)$ and $V_2O_3[100]\|Ni[10\bar{1}]$ would result in a lattice mismatch of only 0.7 %, while for $V_2O_3$ on $Al_2O_3$ substrate a 4 % mismatch can be expected. Reciprocal space mapping (RSM) of the $V_2O_3$ film confirms epitaxial growth of the film. From the RSM and pole figure measurement, Fig. 1b and 1c, respectively, the $(2\bar{1}0)$ and $(01\bar{5})$ diffraction vectors are identified as orthogonal in-plane axes, illustrated in Fig. 1d. Henceforth, all in-plane sample angles ($\phi$) are defined relative to the $(01\bar{5})$ diffraction vector ($\phi=0$). The $V_2O_3$ rhombohedral to monoclinic SPT was measured along the (012), (014) and $(01\bar{4})$ diffractions (not shown) with XRD[24].

The $V_2O_3$/Ni film microstructure obtained from AFM and SEM is shown in Figs. 2a and 2b, respectively. The micrographs identify extended rips and localized voids oriented along the $V_2O_3$ $(01\bar{5})$ diffraction vector (i.e. $\phi=0°$), emphasized by the ellipsoidal shape of the Fourier transformation included in Fig. 2 insets. The rips and voids are approximately 15 nm in width with length ranging from 30 nm to several hundred nanometers. The orientation of the rips is correlated with the morphology of the terraced sapphire substrate, as shown in Fig. 2c. These rips define a microstructure in the bilayer comprised of ribbon-like contiguous regions ≈100 nm in width and >500 nm in length. The Ni control sample of the same thickness grown on a bare r-cut sapphire, shown in Fig. 2d, retains the terrace morphology but does not exhibit any observable rips. This suggests that the growth of the $V_2O_3$ layer underneath the Ni is the cause of the rips. This possibly arises from the Ni film experiencing a particularly strong strain at the terrace boundaries during deposition[25] and the subsequent phase transition, leading to rupture of the film. This would suggest that the structure of the rips may be tuned by designing the terraces or controlling the granular structure of the film, which can be achieved by tuning of the growth parameters.[26]

*Ferromagnetic Resonance*

The oriented microstructural features are expected to introduce a uniaxial shape anisotropy in the plane of the film, which can be probed using angular-dependent FMR measurements. Indeed, the surface map of the FMR spectrum for both the $V_2O_3$/Ni, Fig 3a, and Ni on sapphire, Fig 3b, films show an angular dependence of the resonant field, $H_{Res.}$ identified by the solid lines. Variations in $H_{Res.}$ indicate an anisotropic in-plane effective field. In both samples the minima in $H_{Res.}$ occurs along $\phi=0°$ and 180°, while the maxima occur along $\phi=90°$ and 270°, identifying the magnetic easy and hard axes, respectively. The 180° periodicity indicates the presence of a *uniaxial* anisotropy. For the sapphire/Ni reference sample the anisotropy direction can be correlated to the direction of the terraces in the substrate. In addition, the FMR signal line width of the $V_2O_3$/Ni sample also exhibits an angular dependence, with strong damping parallel to the rips ($\phi=0°$). In comparison, for the Ni on sapphire sample [Fig. 3b], the FMR line width is constant at all angles, indicating that the dissipation processes are angle-independent.

*Magnetometry*

Major hysteresis loops of the $V_2O_3$/Ni sample taken at 295 K, 170 K, 160 K, and 77 K, at angles of -45° → 135° are shown in Figs. 4a-d, respectively. The temperature dependent coercivity $H_C$ and loop squareness $S$ (ratio of remanent over saturation magnetization, $M_R/M_S$) are shown in Fig. 4e and 4g, respectively. Above 170 K the sample has a small coercivity (< 10 mT) which slowly increases with decreasing temperature, consistent with reduced thermal activation. Below the $V_2O_3$ phase transition at 160 K, the Ni coercivity increases to > 20 mT. This has been previously attributed to strain in the Ni film induced by the $V_2O_3$[14, 15, 24, 27]. Here we find a relatively small peak in the temperature-dependent $H_C$ (< 5%, Fig. 4e) presumably due to a large RMS roughness (2.5 nm measured with AFM), as discussed in Ref. 14. Interestingly, the $H_C$ trend shows a strong temperature dependence and very small angular dependence, while $S$ shows a mixed trend. Above the SPT, $S$ depends sensitively on the angle, with a maxima along $\phi=0°$, while below the SPT $S$ exhibits a similarly large value of > 0.6 for all angles. For comparison, the $H_C$ and $S$ trends for the sapphire/Ni sample are included in Figs. 4e and 4f, respectively, exhibiting only a slowly increasing trend with decreasing temperature at all angles, consistent with thermal activation.

Detailed polar contour plots of the angular-dependent coercivity and squareness are shown in Figs. 4f and 4h, respectively. At all temperatures, maxima in $H_C$ and $S$ are observed at $\phi=0°$ [along $(01\bar{5})$] and minima at 90° [along $(2\bar{1}0)$], indicating the magnetic easy and hard axes, respectively, and consistent with the FMR results. Unexpectedly, the uniaxial anisotropy is less pronounced at temperatures below the $V_2O_3$ phase transition at 160 K, as shown in the hard and easy axes coercivities [Fig. 4e] and squareness [Fig. 4g]. The percent difference in $H_C$ and $S$, between the hard and easy axes, can be calculated: $\Delta H_C = \frac{H_C^0 - H_C^{90}}{H_C^0}$ and $\Delta S = \frac{S^0 - S^{90}}{S^0}$. These differences are significantly larger at 295 K, $\Delta H_C$= -73% and $\Delta S$= -79%, than at 77 K, $\Delta H_C$= -27% and $\Delta S$= -14%.

First order reversal curves were measured following published procedures[28, 29, 30, 31], as discussed in Methods. The FORC results provide a mechanism to investigate the quasi-static magnetization reversal characteristics and are shown in Fig. 5 for $\phi=0°$ (top), 45° (middle) and 90° (bottom) at 295 K (left), 160 K (center), and 77 K (right), respectively. A strong variation in the shape of the FORC feature is found as a function of angle and temperature, implying changes in the reversal mechanism. Along the magnetic easy axis, $\phi=0°$, the FORC distribution is oriented along the $H_C^*$ axis for all temperatures. In contrast, along $\phi=90°$ the FORC distribution consists of a left-bending 'boomerang' feature [32], with one arm oriented along the $-H_R$ axis and the other along $-H$, and possessing a negative feature located in the crook of the elbow. FORC distributions measured along $\phi=45°$ are oriented along the $H_C^*$ axis at room temperature, and develop a boomerang structure, with symmetries along $H$ and $H_R$, at lower temperatures. This change in FORC structure implies a temperature and angular dependence of the magnetization reversal mechanism, as discussed below.

**Discussion**

Standard major-loop magnetometry measurements of $V_2O_3$/Ni films show an angular-dependent suppression of $H_C$ and $S$ with minima along the direction orthogonal to the rips ($\phi=90°$), as shown in Figs. 4f and 4h. A similar angular-dependence was observed in the FMR $H_{Res.}$ for both $V_2O_3$/Ni and sapphire/Ni films, with the maxima of $H_{Res.}$ observed along $\phi=90°$. However, only the $V_2O_3$/Ni film showed an angular dependent FMR line width. The key difference in these films is their microstructure, with the $V_2O_3$/Ni film possessing rips, and the sapphire/Ni film possessing only terracing. This suggests that the uniaxial anisotropy seen in the FMR and magnetometry can be attributed to the microstructure, including rips, terracing, and any crystalline texture, while the rips additionally affect the FMR line width. The uniaxial anisotropy can be understood by considering the shape anisotropy: the ribbon-like contiguous regions, bounded by the rips or terrace boundaries, experience a lower magnetostatic interaction when magnetizations are oriented along the ribbon, compared to an orthogonal orientation[33, 34, 35]. The variation in the FMR $H_{Res.}$ identifies the strength of the uniaxial anisotropy as 9.8 mT. Since the room temperature coercivity along the easy axis is only 10 mT, the uniaxial anisotropy is expected to play an important role in the energy landscape. At low temperatures the film's anisotropy increases due to strain from the $V_2O_3$ substrate, increasing the major-loop coercivity while the anisotropy from the microstructure is not expected to change significantly. Thus, the microstructure contributes less to the magnetic behavior below the $V_2O_3$ phase transition, resulting in a suppression of the relative angular dependence.

Using the FORC distributions in Fig. 5, a model of the angle- and temperature-dependent reversal behavior is developed. The film presented in this work is expected to switch by a multi-domain reversal, with each domain geometrically bounded by the rips into approximately 100 nm × 500 nm × 10 nm regions. First, we discuss the FORC distributions measured along $\phi=0°$, which show features oriented along the $H_C^*$ axis, Figs. 5a, 5d and 5g. As discussed in Ref. [30], a FORC feature located on the $H_C^*$ axis identifies symmetric down- and up-switching events ($H = -H_R$). This is common in systems of non-interacting magnetic elements, in which the switching fields are determined only by the local intrinsic anisotropy[36, 37]. A horizontal FORC ridge along the $H_C^*$ axis similarly represents a distribution of local anisotropies. Here we expect domain nucleation to occur at local low anisotropy defect sites within the bounded regions, such as inclusions, grain boundaries, etc. Once the reversal field $H_R$ overcomes the anisotropy of the defect sites, reversal domains are nucleated and then propagate, similar to a non-defective thin film. Following the FORC measurement procedure, the field sweep direction is reversed and $H$ is increased back towards positive saturation. The initial part of the subsequent FORC branch - just after reversing the field sweep direction - is probing $H < -H_R$ (i.e. $H_B < 0$). The fact that the FORC diagrams in Figs. 5a, 5d and 5g show no feature in this region implies that the domain which was nucleated at $H_R$ does not grow or shrink in response to the magnetic field in this field range. The primary FORC features in Figs. 5a, 5d and 5g manifest at $H=-H_R$ (i.e. $H_B=0$). At this field the same defect site which nucleated the negatively-oriented domain from positive saturation now nucleates a new positively-oriented domain, by symmetry arguments. Further increasing $H$, following the FORC branch, probes the region $H > -H_R$ (i.e. $H_B > 0$). The FORC distributions in Figs. 5a, 5d and 5g show no features in this region, indicating that the switching event at $H=-H_R$ saturates the sample. Thus the same defect sites nucleate domains for both field sweep directions, causing a symmetry in the nucleation and annihilation events and giving rise to a feature in the FORC diagram which manifests only on the $H_C^*$ axis ($H_B=0$). This also directly implies that (1) the nucleated domain propagates until it encounters a pinning site which (2) has a de-pinning field larger than the defect site's nucleation field. The width of the $H_C^*$ distribution is reflective of the film being comprised of many such regions, each with a unique nucleating defect site. The width of the $H_C^*$ distribution increases with decreasing temperature, particularly cooling through the SPT, indicating an increase of the pinning potential distribution, likely arising from an inhomogeneous strain landscape. At the coldest temperatures (77 K) the horizontal FORC feature begins to bend, especially at the high and low $H_C^*$ ends. This indicates the onset of a domain growth mechanism, discussed below.

FORC distributions measured at $\phi=90°$, Figs. 5c, 5f and 5i, have extended arms oriented along the $-H_R$ and $-H$ directions, forming a 'boomerang' feature. This feature is observed often in systems which reverse by domain nucleation and growth mechanisms[28, 32, 38, 39]. The 'left-bending' orientation of the boomerang is suggestive of strong exchange interactions (compared to dipolar interactions) within the system[32], as expected for an in-plane film. Similar to the above discussion for the $\phi=0°$ data, once the reversal field $H_R$ overcomes the anisotropy of the defect sites, reversal domains are nucleated which then propagate. Following the FORC measurement procedure, the field sweep direction is reversed, and $H$ is increased back towards positive saturation, initially probing $H < -H_R$ (i.e. $H_B < 0$). The -$H$ arm of the FORC feature resides in this region, indicating that the domains contract in response to the magnetic field before reaching the $H = -H_R$, distinctly different than the $\phi=0°$ measurements. This domain contraction implies that the barrier to domain wall motion is much less for the measurements at $\phi=90°$ compared to the measurements along $\phi=0°$, resulting in an entirely different field evolution. This lack of pinning could indicate that the nucleated domain does not fill the ribbon-like regions after the initial nucleation event, and thus is only weakly pinned. The arm parallel to the $H_R$ axis identifies the progressive negative saturation of the sample and domain re-nucleation from negative saturation[28].

FORCs measured at $\phi=45°$, Figs. 5b, 5e and 5h, show features along $H_C^*$ at room temperature and bending towards a boomerang structure at low temperatures. This explicitly identifies a temperature-dependent transition in the reversal mechanism from localized domain reversal at room temperature to domain growth at low temperatures. This change is possibly due to an increase in strain-induced pinning sites at low temperature due to the coupling of Ni to $V_2O_3$,[14, 15] which prevent the long-range expansion of the domain wall. In this scenario, at high temperatures the magnetic domains propagate over large regions, appearing as switching at single field steps, while at lower temperatures magnetic domains are nucleated and incrementally grow following a Barkhausen-type jumps over multiple field steps.[40] Alternatively, at lower temperatures, the Ni anisotropy increases[14, 15] and correspondingly the domain nucleation field also increases. Thus, at high temperatures, along a FORC branch, domains which are pinned on defect sites at $H_R$ may nucleate a new domains before the domain wall becomes de-pinned, manifesting as reversal in what appears as single-field jumps. At low-temperatures the increased anisotropy may push the nucleation field beyond the de-pinning field, causing an apparent change in the reversal mechanism to an incremental growth mode in which the domain wall jumps between pinning sites.

**Conclusions**

In summary, the magnetization reversal characteristics of $V_2O_3$/Ni bilayer films with a strain-induced uniaxial anisotropy, manifested in the sample microstructures, was investigated. X-ray diffraction and microscopy show rips and voids in the Ni oriented along the $(01\bar{5})$ diffraction vector of the $V_2O_3$. FMR results show these rips attenuate magnon excitation, while strain induced by microstructural terracing causes an effective uniaxial anisotropy field of 9.8 mT, emphasizing the combined contributions from sample microstructure and strain. Standard magnetometry confirmed a uniaxial easy axis, as revealed by the measured coercivity and major loop squareness. The effective anisotropy was demonstrated to be tunable by temperature, with the effect being much weaker below the $V_2O_3$ SPT. First order reversal curve measurements revealed that the anisotropy causes two different reversal mechanisms to exist within the sample, depending on the angle. Along the magnetic easy axis the domains reverse by a defect-induced localized magnetization reversal mechanism, while the hard axis shows reversal by domain nucleation/growth mechanism. FORC measurements performed between the hard and easy axes show a temperature-dependent transition between these reversal mechanisms. This temperature evolution of the magnetic reversal provides insight into Ni microstructure change due to the strain exerted by $V_2O_3$ transition. The results demonstrate an approach to achieve in-plane uniaxial shape anisotropy emergent in the films without patterning or magneto-crystalline considerations, controllable with temperature. Such tunable oriented structures may be relevant to other research fields, including the design of bidirectional phonon conductors for directed thermal transport, or angular dependent plasmon conductors.

**Methods**

Thin films of $V_2O_3$ (100 nm)/Ni (10 nm)/Cu cap (7 nm) and Ni (10 nm)/Nb cap (6 nm) were grown by sputtering in a high vacuum chamber ($P_{Base} \approx 10^{-5}$ Pa) on r-cut (102) sapphire substrates in a 0.53 Pa Ar atmosphere. The $V_2O_3$ was grown at 1000 K substrate temperature from a stoichiometric target and the Ni, Cu, and Nb layers were sputtered at room temperature from elemental targets. The metal-insulator transition in $V_2O_3$ was largest when grown on an r-cut sapphire substrate.[41] X-ray diffraction (XRD) was performed in an inert atmosphere over 90 K - 300 K using Cu-$K_\alpha$ radiation. All in-plane directions are identified by the corresponding (hkl) diffraction vectors of the $V_2O_3$ lattice planes in the high temperature rhombohedral phase using the short hexagonal lattice notation. Surface morphology was investigated by scanning electron microscopy (SEM) and atomic force microscopy (AFM). Magnetic measurements were performed between 77 K and 295 K, using a zero-field cooling sequence, on a vibrating sample magnetometer (VSM) with a liquid nitrogen flow cryostat and the magnetic field was applied in the film plane.

Ferromagnetic resonance (FMR) was measured using an EPR spectrometer with a cylindrical cavity resonator at 9.4 GHz. The sample was placed in the center of the cavity, and mounted on a quartz rod allowing for in-plane orientation of the dc magnetic field. An automatic goniometer allowed for precise rotation with an accuracy of $\Delta\theta = 0.025°$. Alignment of the sample with respect to the ac and dc fields was performed from the highest resonant field while rotating the sample about the out-of-plane direction ($\theta = 0°$) within 0.2°. The dc magnetic field was swept from 0 to 0.9 T at 6 mT/s. The microwave power was kept constant at 1 mW in order to increase the signal-to-noise ratio. The line width ($\Delta H$) and resonance field ($H_{Res.}$) were obtained by fitting the FMR signal to the derivative of a Lorentzian function.

The FORC measurements[42, 43] were performed following procedures discussed previously[28, 29, 30, 31]. Starting from positive saturation, the applied magnetic field was decreased to a reversal field, $H_R$, at which point the field sweep direction was reversed and the magnetization, $M$, is recorded as the applied field, $H$, was increased back to positive saturation. This sequence measures a single FORC branch and is performed for a series of $H_R$ between positive and negative saturation, collecting a family of FORCs. A mixed second order derivative was applied to extract the FORC distribution: $\rho(H, H_R) = -\frac{1}{2}\frac{\partial^2 M(H,H_R)}{\partial H \partial H_R}$. Recognizing that progressing to more-negative $H_R$ probes new down-switching events, and along each FORC branch, increasing $H$ probes up-switching events, a new coordinate system can be defined by the local coercivity, $H_C^* = \frac{H-H_R}{2}$, and bias field, $H_B = \frac{H+H_R}{2}$.

The FORC distribution is a mapping of the switching events within a system. In its most basic application, the reversal mechanism (domain growth, coherent rotation, vortex, etc.) can be uniquely fingerprinted[29, 44]. Furthermore, intricate details of the nanoscale system, including magnetic interactions[30, 32] and intrinsic distributions[36], can be qualitatively, and sometimes quantitatively, identified. One approach which can be used to extract these details from the FORC distribution is to use the $(H, H_R)$ coordinates of features in the FORC distribution and correlate it back to the family of FORCs.

**Acknowledgements**

Work at UCD was supported by the National Science Foundation DMR-1008791 (D.A.G) and ECCS-1611424 (K.L.). The magnetism aspect of this research at UCSD was supported by DOE

Grant No. DE-FG02-87ER-45332 and the oxide related research by AFOSR Grant No. FA9550-12-1-0381. J.T. and I.K.S. thank B. Dodrill for his extensive measurements and interactions. J.T. acknowledges the support from the Fundación Ramón Areces (Spain). D.A.G. acknowledges support from the National Research Council Research Associateship Program. J.G.R. acknowledges the support from FAPA program through Facultad de Ciencias and Vicerrectoria de Investigaciones of Universidad de los Andes, Bogotá Colombia and Colciencias under contracts 120471250659 and 120424054303.

**Author contributions**

The samples were fabricated by J.G.R, J.V. and I.K.S. X-ray measurements and analysis were performed by T.S., J.V. and I.K.S. FMR measurements and microscopy were performed by J.G.R, J.V. and I.K.S. Magnetic characterization and FORC measurements were performed by D.A.G. and K.L. D.A.G., J.V. and K.L. wrote the first draft of the manuscript. All authors contributed to discussions and manuscript revision. J.V. coordinated the project

**Additional information**

Reprints and permissions information is available online at www.nature.com/reprints.

**Competing financial interests**

The authors declare no competing financial interests.

**Figure Captions**

**Figure 1 (color online)** (**a**) High-angle θ-2θ x-ray diffraction pattern of $V_2O_3$ (100 nm)/Ni (10 nm) sample, misaligned by 0.1° to suppress the substrate peaks. The $V_2O_3$ (012), (024), and (036) peaks are identified by *, the Ni (111) identified by †, and sapphire (012) and (024) identified by ‡. (**b**) Reciprocal space map of $V_2O_3$ recorded along the long edge of the substrate. The (012) plane is parallel to the surface at $Q_{//}=0$, while the (104) points 48° to the surface normal. (**c**) Pole figure of the (104) plane showing the epitaxy of $V_2O_3$. The (01$\bar{4}$) lies at 82° with respect to the surface normal. (**d**) Schematic reconstruction of the real space orientation of the $V_2O_3$ crystal planes following from symmetry considerations of the measured planes.

**Figure 2** (**a**) AFM and (**b**) SEM image of the $V_2O_3$/Ni film, (**c**) AFM of the sapphire substrate and (**d**) AFM of Ni on sapphire. The image - supported by Fourier transform (insert) - identifies many rips oriented along the (01$\bar{5}$) diffraction vector. Scale bars indicate 1 μm.

**Figure 3 (color online)** Angular-dependent ferromagnetic resonance spectrum of (**a**) $V_2O_3$/Ni and (**b**) Ni on sapphire. Contrast indicates the differential absorption. The peak in the absorption for each angle is identified by the $H_{Res.}$ lines. The y-axis identifies the angle $\phi$ between the applied magnetic field and the (01$\bar{5}$) diffraction vector. $H_{Res.}$ is indicated by the solid curve.

**Figure 4 (color online)** Major hysteresis loops of $V_2O_3$/Ni taken at (**a**) 295 K, (**b**) 170 K, (**c**) 160 K, and (**d**) 77 K measured at 0°, 30°, 45°, 60°, and 90° (red to blue). Trends for coercivity $H_C(T,\phi)$ and squareness $S(T,\phi)$ (**g, h**) from (**a-d**) are collated (**e, f**) and (**g, h**), respectively. Results of the reference $Al_2O_3$/Ni sample are included as dashed curves in (e) and (f) for comparison.

**Figure 5 (color online)** FORC distributions measured at (top) 0° [along the $(01\bar{5})$ diffraction vector], (middle) 45°, and (bottom) 90° [along the $(2\bar{1}0)$ diffraction vector], at (left) 295 K, (center) 160 K, and (right) 77 K. The color range in each contour extends between the maximal and minimal value, with the background, away from the feature, identifying zero. Arrows plotted in each panel indicate the $H$ and $H_R$ coordinate axes.

# Figures

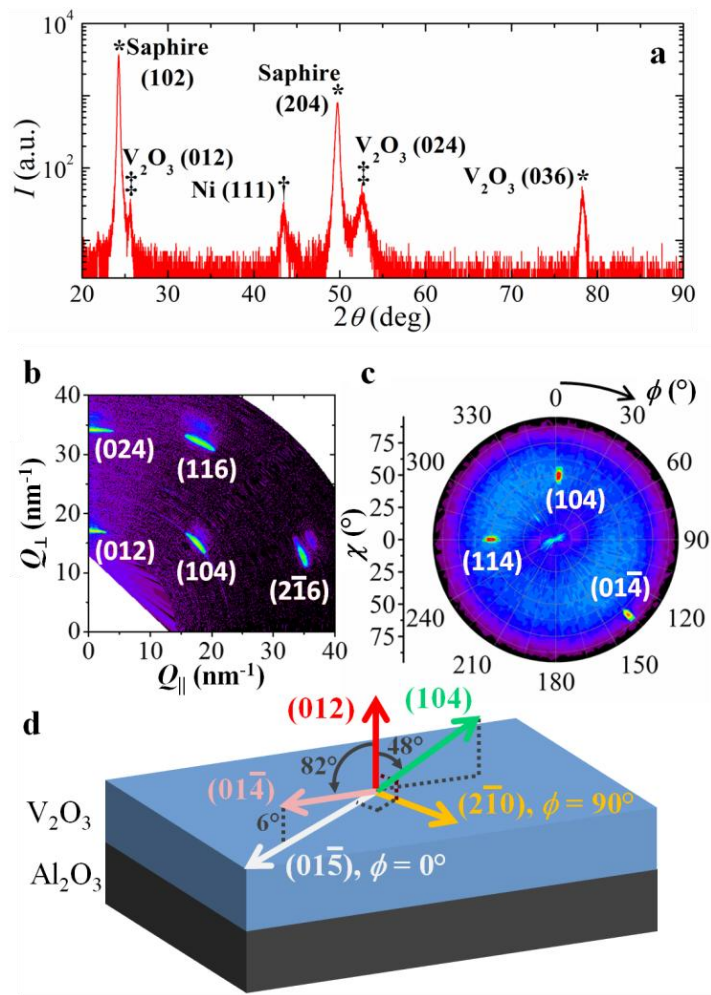

Fig. 1

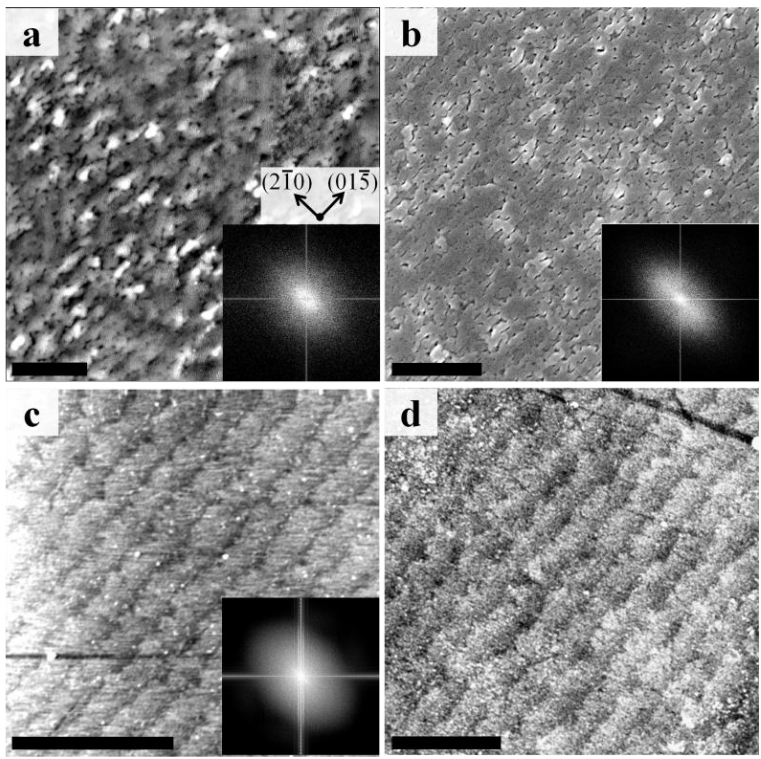

Fig. 2

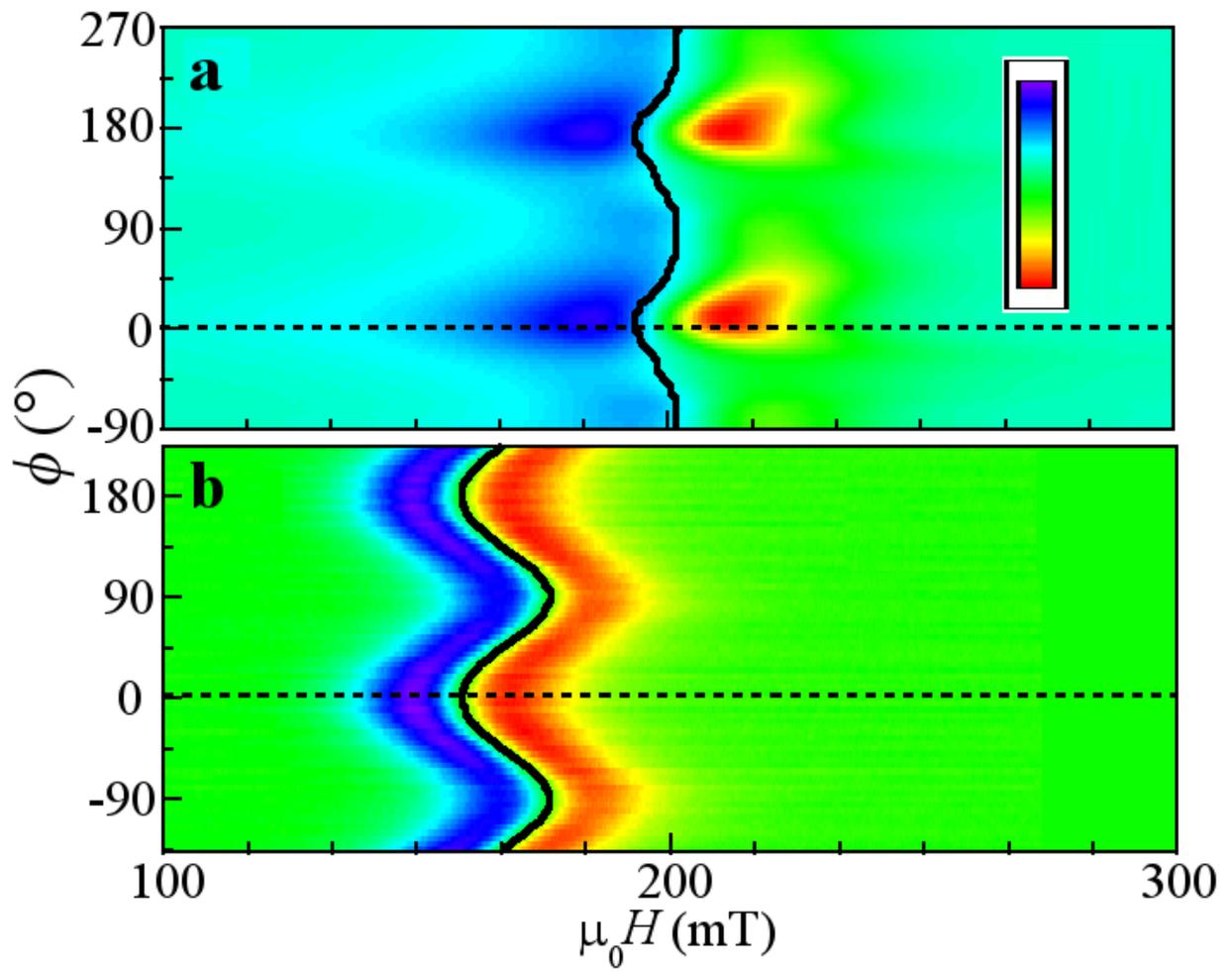

**Fig. 3**

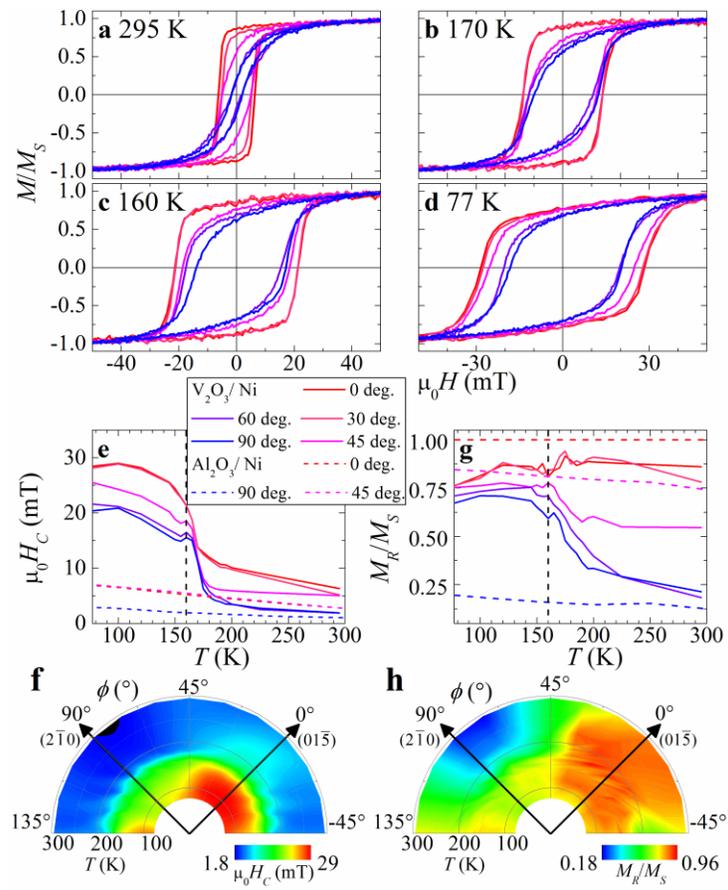

**Fig. 4**

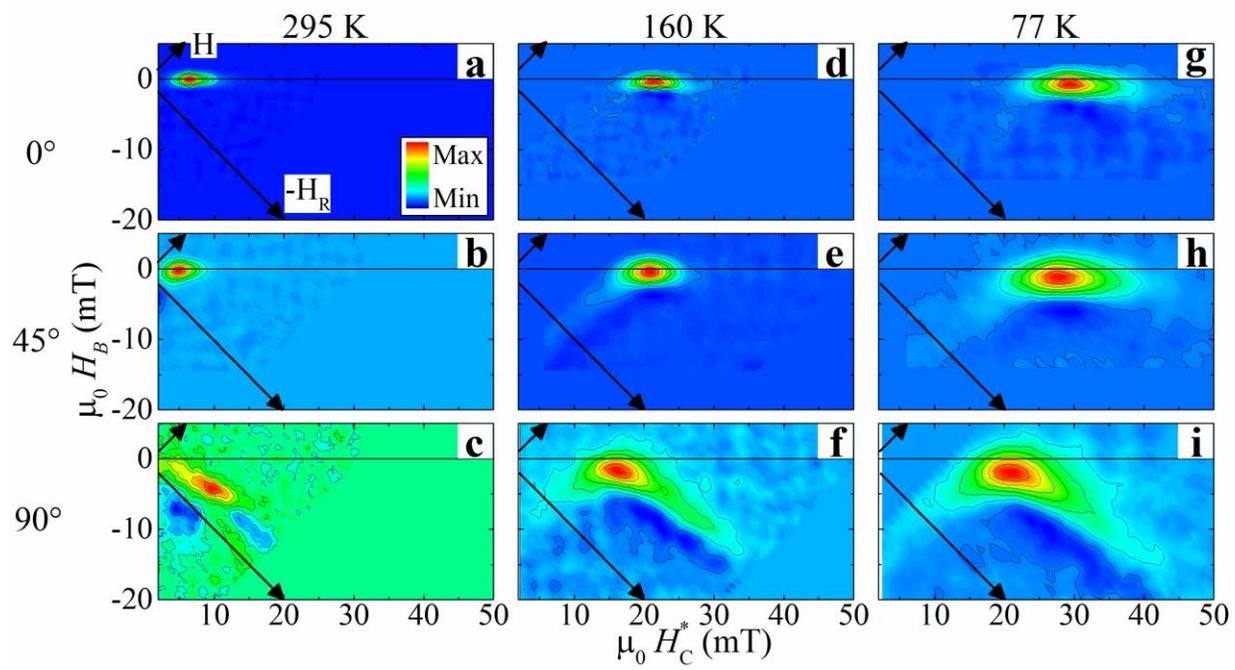

**Fig. 5**